\documentclass{article}

\usepackage{arxiv}

\usepackage[utf8]{inputenc} 
\usepackage[T1]{fontenc}    
\usepackage{hyperref}       
\usepackage{url}            
\usepackage{booktabs}       
\usepackage{amsfonts}       
\usepackage{nicefrac}       
\usepackage{microtype}      
\usepackage{lipsum}
\usepackage{graphicx}
\usepackage{multirow}
\usepackage{epstopdf}
\usepackage{verbatim}
\usepackage{subfig}

\usepackage{cite}   

\title{Experimental methods for the study of standing waves in strings}

\author{
  K. L. Cristiano \\
  Facultad de Ciencias, Escuela de Física\\
  Universidad Industrial de Santander (UIS)\\
  Colombia, Santander, Bucaramanga \\
  \texttt{karen.cristiano@saberuis.edu.co} \\
  \And
   D. A. Triana \\
	Facultad de Ciencias, Escuela de Física\\
	Universidad Industrial de Santander (UIS)\\
	Colombia, Santander, Bucaramanga \\
   \texttt{ dantrica@saber.uis.edu.co } \\
   \And
   R. Ortiz \\
   Departamento De Matemáticas Y Ciencias Naturales\\
   Universidad Autónoma de Bucaramanga (UNAB)\\
   Colombia, Santander, Bucaramanga\\
   \texttt{rortiz579@unab.edu.co} \\
   \And
      A. F. Estupiñán \\
      Departamento de Ciencias Básicas y Humanas\\
      Universidad de Investigación y Desarrollo (UDI)\\
      Colombia, Santander, Bucaramanga\\
      \texttt{aestupinan4@udi.edu.co} \\
}

\begin{document}
\maketitle


\begin{abstract}

Several cases of the explanation of the phenomenon of standing waves in strings, there are few experimental measurement tools when demonstrating this phenomenon in a classroom, it is for this reason that we have implemented different forms to show how we can experimentally demonstrate the wave behavior of string vibration, where variations in the length, frequency and tension of the string have been made, with the purpose of showing this phenomenon more generally and clearly for students.

In this work, we present the step-by-step the implementation of two experimental procedures of laboratory concerned with the study of wave strings, seeking as a primary objective to show students the experimental way of obtaining and measuring the number of bellies in a string in addition of being able to calculate both fundamentally and experimentally the fundamental frequency of each string by varying both its length and its tension using the resonance frequency.

Finally, we present the respective experimental errors, for each of the assemblies and methods carried out in the laboratory to perform the measurements shown and registered in this work.

\end{abstract}

\keywords{Fundamental frequency \and Resonance frequency \and Bellies \and Nodes \and Waves on strings.}


\section{Introduction}

Considering the importance in the course of Physics III that has the phenomenon of waves in strings and tubes \cite{alex_1}, we want to show in a clearer and more practical way this phenomenon, for example in the references \cite{c1,c2,c3,c4}. We propose in this work, the possibility of verifying both experimentally and analytically unknown string variables as for example: the volumetric and linear mass density which are intrinsic data of the material and we have calculated them indirectly through an experimental procedure shown in this document.

We, the authors of this article, mainly want in this research work to make two experimental arrangements in which, for the first case we show a method, in which the tension of the string has been modified and the length has been left constant of the same and for the second part of this first case we have varied the length of the string and keep constant the tension applied to the string.

In the second case, we have used a strobe light, with which we want to use the definition of resonance to be able to measure the frequency of vibration of the string directly using a rotary motor, to which the input voltage is varied, which It allows us to have a control of the vibration frequency of the same \cite{sears,serway}.

After performing these two experimental methods, we have obtained the results, in order to indirectly find the linear mass density of each of the strings used in these two experimental arrangements. In order to show students two different and alternative methods to calculate the linear mass density of the nylon strings used to perform these experiments.

This paper is organized as follows: Section \ref{analítico}, describes the Theoretical study corresponding to the calculates for the theory about of the waves in strings. In Section \ref{experimental}, shown the two experimental method procedure for the data takes. In Section \ref{results}, we show the experimental results and we compared the results obtained with this two methods presented in this article.


\section{Theoretical framework}
\label{analítico}

When we want to model a wave phenomenon in physics, we must start from the second order differential equation of the propagation of a wave \cite{zill}, which is posed in Equation (\ref{eq1}) 

\begin{equation}
    \frac{\partial \psi^2}{\partial t^2} = \left( v^2 \right ) \frac{\partial \psi^2}{\partial x^2},
    \label{eq1}
\end{equation}

If, now we propose the function that models a simple harmonic oscillatory motion (See Equation (\ref{eq2})), which is obtained as a solution to equation (\ref{eq1}).

\begin{equation}
    \psi (x,t) = A \cdot \sin (kx - \omega t)
    \label{eq2}
\end{equation}

Now, taking into account the wave propagation equation in the string, we can write the wave equation, based on the Tension $T$ applied to it and the linear mass density (See Equation (\ref{eq3})),

\begin{equation}
 \frac{\partial \psi^2}{\partial t^2} = \left( \frac{T}{\mu} \right ) \frac{\partial \psi^2}{\partial x^2},    
 \label{eq3}
\end{equation}

Comparing the equation of motion (1) with (3), we have the value of the propagation velocity of the wave which is:

\begin{equation}
    v^2 = \frac{T}{\mu}, \quad  
\end{equation}

\begin{figure}[h]
	\centering
\includegraphics[width=0.85\textwidth]{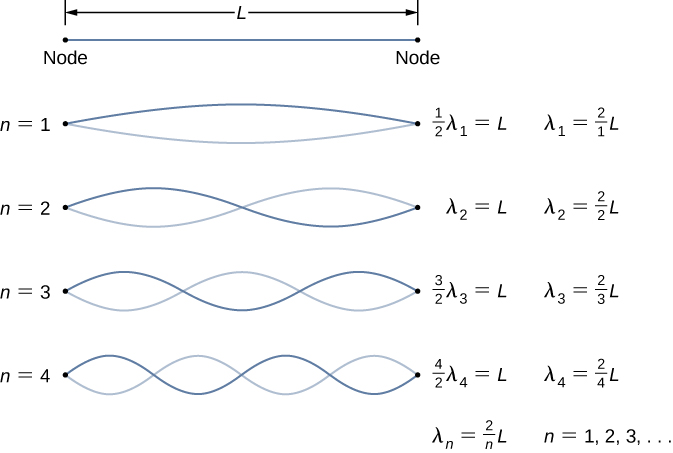}
	\caption{Behaviour of a stationary wave, where we show the different bellies and nodes presents in the string. This image was extracted from \cite{sears}.}
	\label{fig_1}
\end{figure}

The phenomenon physics of our system to study, consists in the behaviour of a stationary wave in a string fixed in its two extremes, how we shown in the Figure \ref{fig_1}.

It is important to note that in Figure \ref{fig_1}, the value of $n$ corresponds to the number of bellies reached as a function of the wavelength $\lambda$ of the string. The equations, which govern the vibrational behavior of a string attached to its two ends, are given by the equations (\ref{eqc_1}) and (\ref{eqc_2}).

\begin{equation}
    v = \lambda_n \cdot f = \frac{2}{n}L \cdot f, 
\label{eqc_1}
\end{equation}

\begin{equation}
    v = \sqrt{\frac{T}{\mu}}.
    \label{eqc_2}
\end{equation}

Where $T$ is the tension applied to the string, lambda is the wavelength, f is the frequency of vibration, $v$ the speed of propagation of the wave, and $\mu$ is the density of the line's mass of the string for different values of $n$ bellies \cite{tipler}. By relating equation (\ref{eqc_1}) and equation (\ref{eqc_2}), we can obtain Equation (\ref{eqc_3}), which directly relates the applied tension of the string according to the frequency variation of the wave present in the string.

\begin{equation}
T = \frac{4 L^2 f^2 \mu}{n^2}, \quad \omega_n = \sqrt{\frac{T}{\mu}} \cdot \frac{n \cdot \pi}{L}
\label{eqc_3}
\end{equation}

\section{Experimental study}
\label{experimental}

\subsection{Method 1}

\begin{figure}[h]
 \centering
    \includegraphics[width=0.7\textwidth]{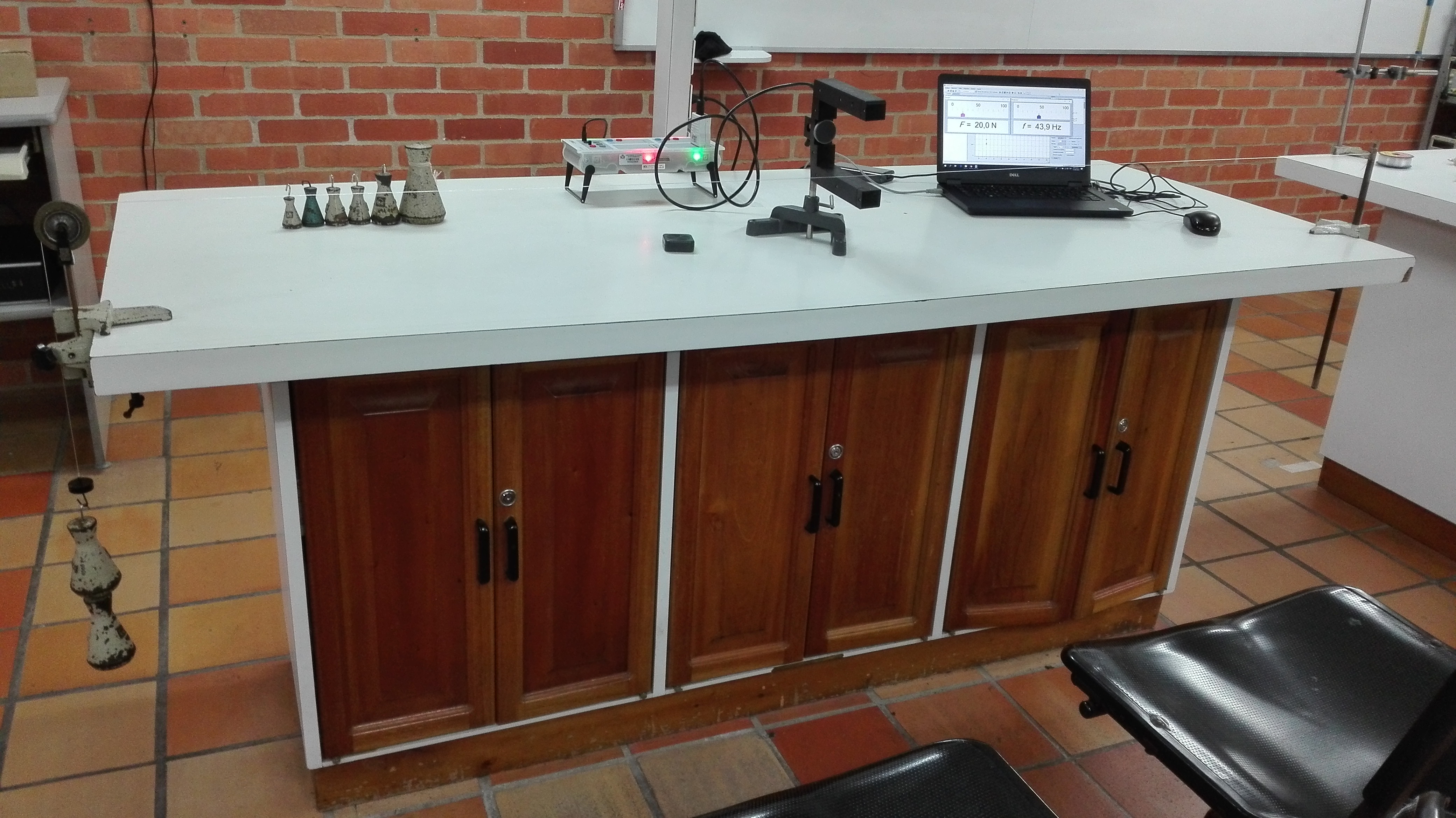}
 \caption{Experimental arrangement for the first method, where the main materials to be used are shown.}
 \label{figure_montaje}
\end{figure}

To carry out the experimental study of this work, we began to elaborate the assembly shown in Figure \ref{figure_montaje}, corresponding to method 1.

In Figure \ref{figure_montaje}, we can see that in this case, the experimental assembly that was performed with the purpose of measuring the frequency of vibration of the string, we use the Software shown in Figure \ref{figure_montaje} that we have implemented, we have classified the data of as follows.

For the first part of method 1, we have varied the length of the string and keep the tension of the string constant as shown in Table \ref{table_1}.

\begin{table}[h]
\centering
\begin{tabular}{|c|c|c|c|c|}
\hline
\textbf{L {[}m{]}} & \textbf{$f_1$ {[}$Hz${]}} & \textbf{$f_2$ {[}$Hz${]}} & \textbf{$f_3$ {[}$Hz${]}} & \textbf{$f_{average}$ {[}$Hz${]}} \\ \hline
0.2                & 701.1                     & 699.2                     & 704.2                     & 701.5                             \\ \hline
0.4                & 345.6                     & 348                       & 357                       & 350.2                             \\ \hline
0.6                & 239                       & 236.5                     & 235.8                     & 270.43                            \\ \hline
0.8                & 178                       & 178.6                     & 177.5                     & 178.03                            \\ \hline
1.0                & 142.5                     & 142.5                     & 142.2                     & 142.23                            \\ \hline
1.2                & 118.7                     & 118.6                     & 118.7                     & 118.67                            \\ \hline
1.4                & 101.7                     & 100.8                     & 101.1                     & 101.2                             \\ \hline
1.6                & 87.9                      & 88.7                      & 88.1                      & 88.23                             \\ \hline
1.8                & 78.1                      & 78.1                      & 77.9                      & 78.03                             \\ \hline
2.0                & 70.8                      & 69.8                      & 70.2                      & 70.26                             \\ \hline
2.23               & 62.92                     & 63.3                      & 62.7                      & 62.97                             \\ \hline
\end{tabular}
\vspace{0.2cm}
\caption{Length variation data keeping the string tension constant for a value of T = 39.66 N.}
\label{table_1}
\end{table}

In the second part of experimental method 1, we have kept the length of the string constant and now we will vary the tension applied to the string as shown in Table \ref{table_2}.

\begin{table}[h]
\centering
\begin{tabular}{|c|c|c|c|c|c|}
\hline
\textbf{Mass  {[}kg{]}} & \textbf{Tension {[}$N${]}} & \textbf{$f_1$ {[}$Hz${]}} & \textbf{$f_2$ {[}$Hz${]}} & \textbf{$f_3$ {[}$Hz${]}} & \textbf{$f_{average}$ {[}$Hz${]}} \\ \hline
2.0433                  & 20.044                      & 43.7                      & 44.3                      & 44.4                      & 44.13                             \\ \hline
2.54                    & 24.917                      & 48.9                      & 49.6                      & 49.9                      & 49.47                             \\ \hline
3.0433                  & 29.854                      & 53.6                      & 53.9                      & 54.5                      & 54.00                             \\ \hline
3.54                    & 34.727                      & 58.3                      & 58.9                      & 58.6                      & 58.6                              \\ \hline
4.043                   & 39.662                      & 62.5                      & 62.3                      & 62.5                      & 62.43                             \\ \hline
4.54                    & 44.537                      & 66.2                      & 66.4                      & 67.0                      & 66.53                             \\ \hline
4.94                    & 48.461                      & 69.6                      & 69.0                      & 69.5                      & 69.37                             \\ \hline
5.2404                  & 51.408                      & 71.9                      & 71.5                      & 71.8                      & 71.73                             \\ \hline
\end{tabular}
\vspace{0.2cm}
\caption{Taking experimental data, for the case in which the length of the string is kept constant $L$ $=2.23$ $m$, and the tension applied to it is varied, varying the masses placed at one end of the string (See Figure \ref{figure_montaje}).}
\label{table_2}
\end{table}

It is important to remember that in tables \ref{table_1} and \ref{table_2} , the experimental data for the same string have been taken, which has the following characteristics: radius $R$ $=$ $7.75 \times 10^{-4}$ $m$, whose linear mass density is equal to $50.33$ $\times 10^{-5}$ $kg/m$ (corresponding to nylon) and its volumetric mass density is $1.14$ $\times 10^{3}$ $kg/m^{3}$.

\subsection{Method 2}

\begin{figure}[h]
 \centering
\includegraphics[width=0.8\textwidth]{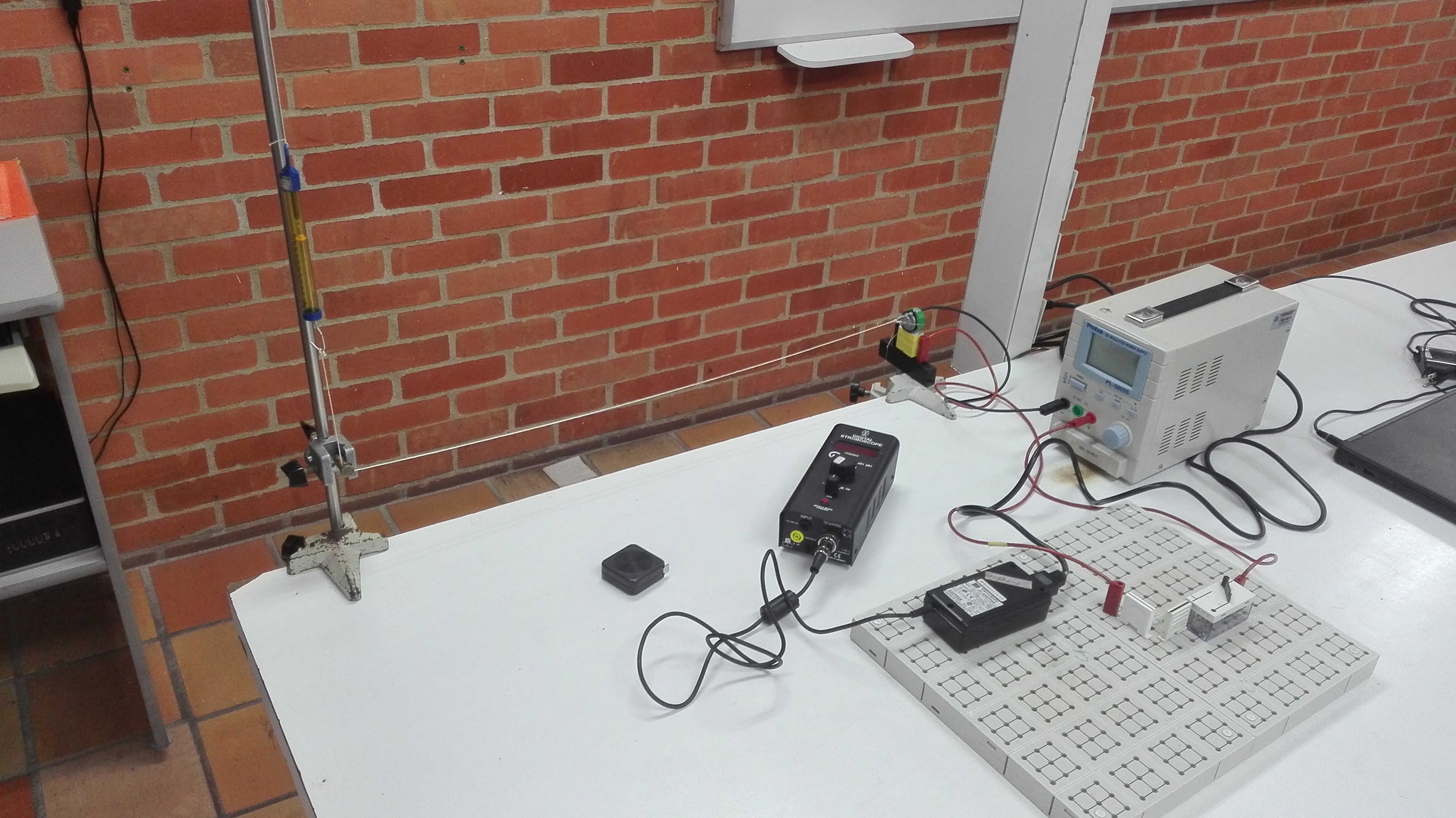}
  \caption{Experimental arrangement for the second method, where the main materials to be used are shown.}
 \label{fig_3}
\end{figure}

The experimental setup that we have done for this method, we show in Figure \ref{fig_3}.

In method 2, we have used a strobe light, with which we can know the frequency of turning it on and off, in addition in Figure \ref{fig_3} the location of a rotation motor with variable rotation frequency is observed at which a variable voltage source has been connected to it as can be seen in Figure \ref{fig_3}.

We collect the data shown in Table \ref{table_3}, where we have used the physical resonance phenomenon in which we take the frequency that appears in the strobe light instrument; just in the instant when it emits the light at the same frequency that the motor that makes the string vibrate.

\begin{table}[h]
\centering
\begin{tabular}{|c|c|c|c|c|}
\hline
\textbf{Number of bellies $n$} & \textbf{$f_1$ {[}$Hz${]}} & \textbf{$f_2$ {[}$Hz${]}} & \textbf{$f_3$ {[}$Hz${]}} & \textbf{$f_{average}$ {[}$Hz${]}} \\ \hline
1                              & 7.43                     & 7.42                     & 7.40                        & 7.42                              \\ \hline
2                              & 14.87                     & 14.85                     & 14.89                     & 14.87                             \\ \hline
3                              & 22.31                     & 22.30                     & 22.29                     & 22.30                             \\ \hline
4                              & 28.15                     & 28.98                     & 28.95                     & 28.69                             \\ \hline
5                              & 40.07                     & 39.51                     & 38.26                     & 39.28                             \\ \hline
\end{tabular}
\vspace{0.2cm}
\caption{Data taken for a constant Tension of $0.05$ $N$, for the first five bellies.}
\label{table_3}
\end{table}

Next, we take the experimental data of the period of oscillation of the string, varying the tension applied to it, keeping its length constant. The data obtained for this part, were recorded in table \ref{table_4}.

\begin{table}[h]
\centering
\begin{tabular}{|c|c|c|c|c|c|}
\hline
\textbf{Tension [N]} & \textbf{$T_1$ [s]} & \textbf{$T_2$ [s]} & \textbf{$T_3$ [s]} & \textbf{$T_{average}$ [s]} & \textbf{$f_{average}$[Hz]} \\ \hline
2                        & 0.176                  & 0.175                  & 0.176                  & 0.175                          & 5.692                          \\ \hline
4                        & 0.142                  & 0.141                  & 0.142                  & 0.141                          & 7.058                          \\ \hline
6                        & 0.118                  & 0.118                  & 0.117                  & 0.117                          & 8.498                          \\ \hline
8                        & 0.101                  & 0.101                  & 0.102                  & 0.101                          & 9.868                          \\ \hline
10                       & 0.09                   & 0.09                   & 0.09                   & 0.090                          & 11.111                         \\ \hline
12                       & 0.082                  & 0.082                  & 0.082                  & 0.082                          & 12.195                         \\ \hline
14                       & 0.076                  & 0.077                  & 0.076                  & 0.076                          & 13.100                         \\ \hline
16                       & 0.072                  & 0.072                  & 0.072                  & 0.072                          & 13.888                         \\ \hline
\end{tabular}
\vspace{0.2cm}
\caption{Data taken for a constant length of $2.5$ $m$, for the first bellies, when we variate the tension of string.}
\label{table_4}
\end{table}


\section{Results}
\label{results}

Starting from the data in Table \ref{table_1}, we can experimentally find the linear mass density of the string, if we plot on the Y-axis the square of the length of the string and on the X-axis we graph the inverse of the square of the frequency, taking into account that for this case, we have kept constant the tension applied to the string.

\begin{figure}[h]
 \centering
\includegraphics[width=1.0\textwidth]{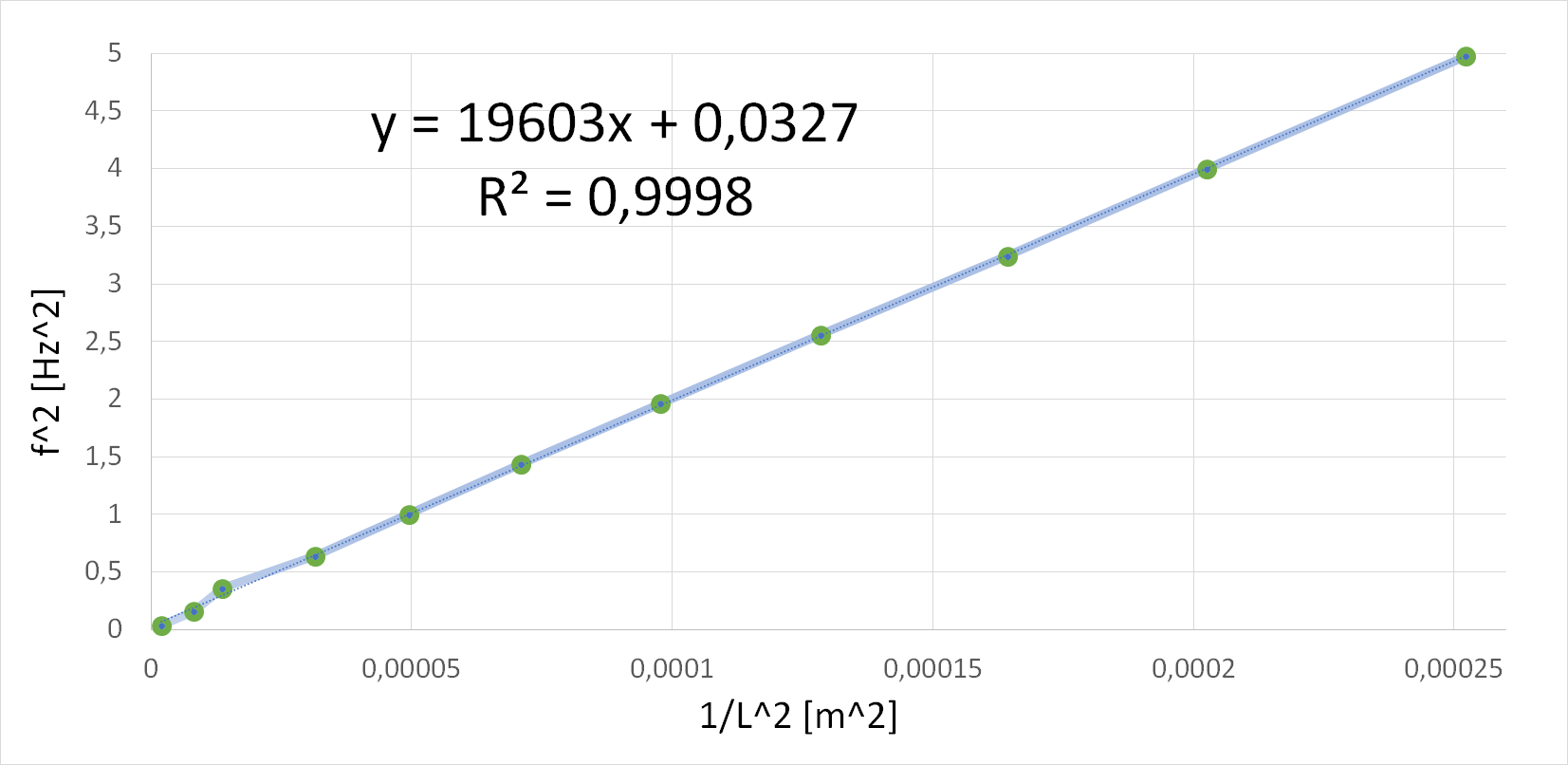}
  \caption{Graph of the relation of the frequency of oscillation of the string in function of the variation of the length, keeping constant the tension $T$ $=$ 3$9.66$ $N$.}
 \label{fig_4}
\end{figure}

Starting from the data in Table \ref{table_1}, we can experimentally find the linear mass density of the string, if we plot on the Y-axis the square of the length of the string and on the X-axis we graph the inverse of the square of the frequency (See Figure \ref{fig_4}), taking into account that for this case, we have kept constant the tension applied to the string. We obtained Equation (\ref{eq_8}), from Equation (\ref{eqc_3}).

\begin{equation}
f^2 = \frac{T n^2}{4 L^2 \mu}
    \label{eq_8}
\end{equation}

\begin{figure}[h]
 \centering
\includegraphics[width=1.0\textwidth]{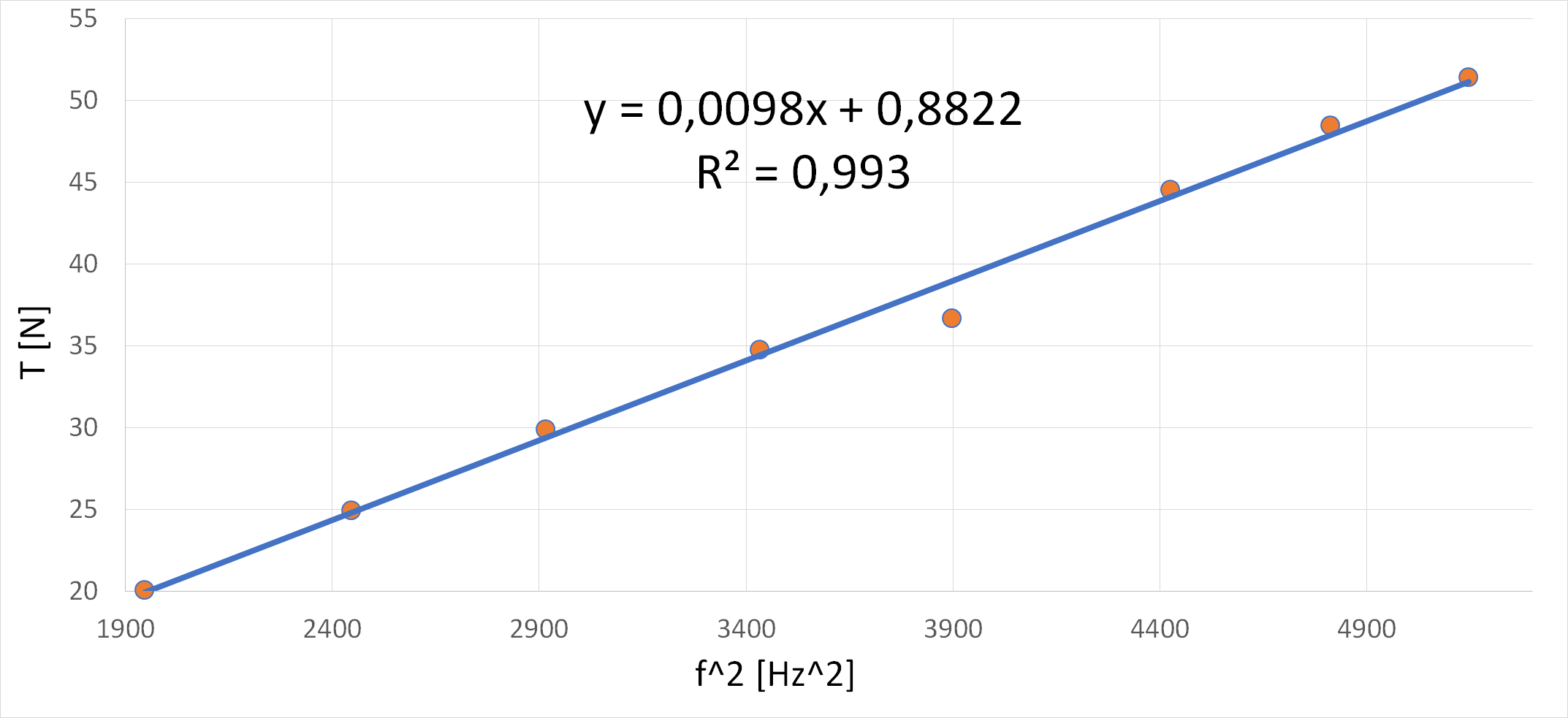}
  \caption{Graph of the tension in the string as a function of the frequency generated in the vibration of the string, for a constant length of $L$ $=$ $2.23$ $m$.}
 \label{fig_5}
\end{figure}

Using Data Table \ref{table_2} and the Equation (\ref{eqc_3}), we can relate the tension of the string according to the frequency obtained, keeping constant the value of the tension applied on the string. We can compare the frequency variation as the tension in the string changes, based on the data taken in Table \ref{table_2}, we can obtain a graphical relationship of these results shown in Figure (\ref{fig_5}).

From the Equation (\ref{eq_8}), if we clear the frequency of oscillation of the string, depending on the number of nodes generated in the string, we can reach Equation (\ref{eq_9}).

\begin{equation}
n^2 = \frac{4 L^2 \mu f^2}{T}
\label{eq_9}
\end{equation}

\begin{figure}[h]
 \centering
\includegraphics[width=1.0\textwidth]{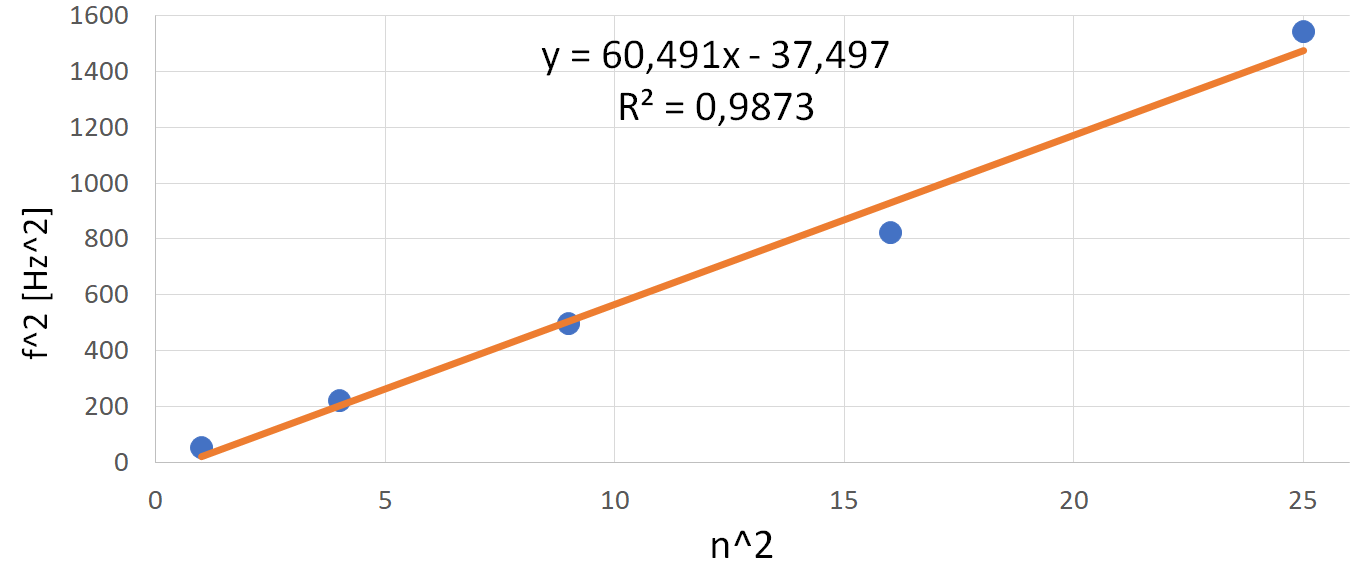}
  \caption{Graph of the number of bellies in the string as a function of the frequency generated in the vibration of the nylon's string, for a constant length of $L$ $=$ $0.71$ $m$.}
 \label{fig_6}
\end{figure}

In the implementation of the second experimental method, we have obtained the data of the frequency and the number of bellies, if we graph the frequency according to the number of bellies, we can experimentally calculate the linear mass density $\mu$ of it (See Figure \ref{fig_6}).

\begin{figure}[h]
 \centering
\includegraphics[width=0.95\textwidth]{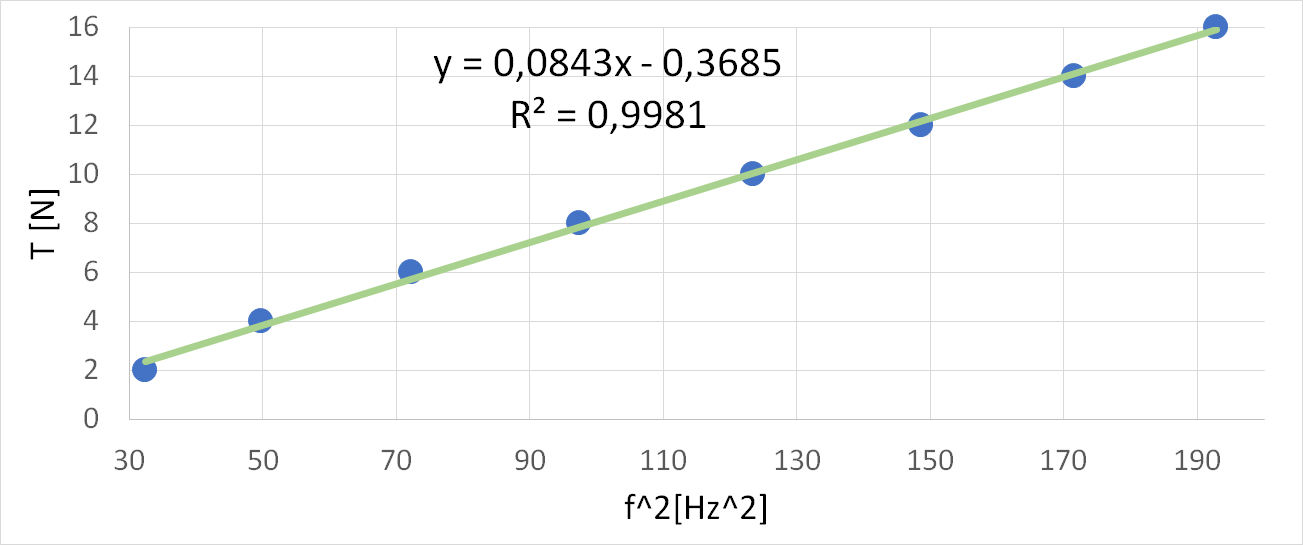}
  \caption{Graph of the voltage as a function of the frequency, for a constant length of $L$ $=$ $2.21$ $m$.}
 \label{fig_7}
\end{figure}

Using now the data in Table \ref{table_4}, we can graph the string tension as a function of the frequency for the second method, we obey the relationship shown in Figure \ref{fig_7}.

After performing this study, the following results were obtained for the calculation of the linear mass density $\mu$ of the strings using both methods, (See Table \ref{table_5}).

\begin{table}[h]
\centering
\begin{tabular}{|c|c|c|c|c|}
\hline
\textbf{linear mass density}        & \multicolumn{2}{c|}{\textbf{Method 1}}              & \multicolumn{2}{c|}{\textbf{Method 2}}             \\ \hline
$\mu$ (Theoretical) {[}$kg/m^3${]}  & 5.057 $\times$ $10^{-4}$ & 5.057 $\times$ $10^{-4}$ & 4.48 $\times$ $10^{-4}$ & 4.48 $\times$ $10^{-4}$  \\ \hline
$\mu$ (Experimental) {[}$kg/m^3${]} & 5.035 $\times$ $10^{-4}$ & 4.926 $\times$ $10^{-4}$ & 4.1 $\times$ $10^{-4}$  & 4.315 $\times$ $10^{-4}$ \\ \hline
\end{tabular}
\vspace{0.2cm}
\caption{Experimental results of the linear mass density $\mu$ of the strings, obtained from the data taken in the Physics laboratory.}
\label{table_5}
\end{table}

Finally, in table \ref{table_6}, we have recorded the experimental errors obtained in our research work. Where we can compare the results obtained in this experiment, for two different experimental methods taking into account that these have been performed, in order to obtain the value of linear mass density from indirect methods, which have been shown in this article.

\begin{table}[h]
\centering
\begin{tabular}{|c|c|c|c|c|}
\hline
\textbf{Errors}                        & \multicolumn{2}{c|}{\textbf{Method 1}}            & \multicolumn{2}{c|}{\textbf{Method 2}} \\ \hline
Absolute Error of $\mu$ {[}$kg/m^3${]} & 2.2 $\times$ $10^{-6}$ & 5.057 $\times$ $10^{-4}$ & 0.38               & 0.16              \\ \hline
\% Relative error of $\mu$             & 0.43                   & 2.59                     & 8.48               & 3.68              \\ \hline
\end{tabular}
\vspace{0.2cm}
\caption{Experimental errors in the calculation of linear mass density $\mu$ of the strings.}
\label{table_6}
\end{table}


\section{Conclusions}

In this work, we have been able to implement the measurement of linear mass density through two experimental methods in a physics laboratory; of two strings of different macroscopic characteristics, in which the implementation and generation of new laboratory materials has been carried out, in order to present two innovative proposals to measure the value of $\mu$ using the wave phenomenon of vibration of the strings.

Finally, in this article we have obtained very reliable results from our measurements, where we were able to register experimental errors below 10\%, which guarantees us the reliability of the data taken and the certification of the good results obtained in these two experiments. In addition, we were able to implement the measurement of a physical magnitude of the string ($\mu$) using two independent experimental methods.


\section*{Acknowledgments}

The authors would like to thank the Universidad Autónoma de Bucaramanga (UNAB), for lend us the installations and materials for carry to this experiment with which the analytical model presented in this paper could be validated.

\end{document}